\documentclass{article}
\usepackage{spconf,amsmath,graphicx,amssymb,bm}
\usepackage{enumitem}% http://ctan.org/pkg/enumitem
\usepackage{mathtools}
\usepackage{siunitx}
\usepackage{cleveref}
\Crefname{section}{Sec.}{Secs}  % https://tex.stackexchange.com/a/386310/148912

\usepackage[acronym]{glossaries}
\usepackage{balance,cite}
\usepackage{comment}

\usepackage{tabularx, etoolbox, booktabs}
\usepackage{multirow} % Allows rowspan
\newcolumntype{H}{>{\setbox0=\hbox\bgroup}c<{\egroup}@{}}  % https://texblog.org/2014/10/24/removinghiding-a-column-in-a-latex-table/
\usepackage{subcaption}
\usepackage{color}
\usepackage{algorithm}
\usepackage{algpseudocode}
\usepackage{color}

\title{Tight integration of neural- and clustering-based diarization\\
       through deep unfolding of infinite Gaussian mixture model \vspace{-2mm}
}
\name{Keisuke Kinoshita, Marc Delcroix, Tomoharu Iwata\vspace{-2mm}}
\address{NTT Corporation, Japan \vspace{-3mm}}

\begin{document}
\ninept
\maketitle
\begin{abstract}
\vspace{-1mm}
Speaker diarization has been investigated extensively 
as an important central task for meeting analysis.
Recent trend shows that integration of end-to-end neural (EEND)-
and clustering-based diarization is a promising approach to handle
realistic conversational data containing overlapped speech with an arbitrarily large number of speakers, and achieved state-of-the-art results on various tasks.
However, the approaches proposed so far have not realized {\it tight} integration yet, 
because the clustering employed therein was not optimal in any sense for clustering the speaker embeddings
estimated by the EEND module.
To address this problem, this paper introduces a {\it trainable} clustering algorithm
into the integration framework, 
by deep-unfolding a non-parametric Bayesian model called the infinite Gaussian mixture model (iGMM).
Specifically, the speaker embeddings are optimized during training such that it better fits 
iGMM clustering, based on a novel clustering loss based on Adjusted Rand Index (ARI).
Experimental results based on CALLHOME data show that 
the proposed approach outperforms the conventional approach
in terms of diarization error rate (DER), 
especially by substantially reducing speaker confusion errors,
that indeed reflects the effectiveness of the proposed iGMM integration. 
\end{abstract}
\vspace{-1mm}
\section{Introduction}
\vspace{-2mm}
\label{sec:intro}
Automatic meeting/conversation analysis is one of the essential technologies required 
for realizing futuristic speech applications such as communication agents that can follow, respond to, and facilitate our conversations. 
% Speaker diarization plays a central part for such meeting analysis systems \cite{Diarization_review, DIHARD_data, AMI_data}.
As an important central task for the meeting analysis, speaker diarization has been extensively studied \cite{Diarization_review, DIHARD_data, AMI_data}.

Current competitive diarization approaches can be categorized into three types; 
speaker embedding clustering-based approaches \cite{x-vector,Diarization_review,DIHARD_JHU,DIHARD_BUT}, 
neural end-to-end diarization (EEND) approaches \cite{Fujita_IS2019,Fujita_ASRU2019,Horiguchi2020_EDA_EEND}, 
and combination/integration of the former two approaches
\cite{EEND-vector-clustering_ICASSP2021,EEND-vector-clustering_Interspeech2021, Horiguchi_ASRU2021, coria2021overlapaware}.
% Among them, the clustering-based approach has the longest research history, and
% is based on clustering of speaker embeddings such as i-vectors \cite{i-vector} and x-vectors \cite{x-vector}.
% They 
The speaker embedding clustering-based approaches 
first segment a recording into short homogeneous chunks 
and compute speaker embeddings such as x-vectors \cite{x-vector} for each chunk assuming that only one speaker 
is active in each chunk.
Then, the speaker embeddings are clustered to regroup segments belonging to the same speakers and obtain the diarization results. 
While these methods can cope with very challenging scenarios \cite{DIHARD_JHU,DIHARD_BUT}
and work with an arbitrarily large number of speakers,
there is a clear disadvantage that they cannot handle overlapped speech.
% i.e., time segments where more than one person is speaking, 
% because of the way of extracting speaker embeddings. 

The second category of diarization approaches, EEND, was recently developed \cite{Fujita_IS2019,Fujita_ASRU2019,Horiguchi2020_EDA_EEND} 
to specifically address the overlapped speech problem.
Similarly to the neural source separation \cite{Kolbaek2017,RSAN},
a Neural Network (NN) receives frame-level spectral features 
and directly outputs a frame-level speaker activity for each speaker, 
no matter whether the input signal contains overlapped speech or not.
While the system is simple and has started outperforming the conventional clustering-based algorithms \cite{Fujita_ASRU2019,Horiguchi2020_EDA_EEND},
it still has difficulty in generalizing to recordings containing a large number of speakers \cite{Horiguchi2020_EDA_EEND}.

To this end, the third category of diarization approaches,
integration of the EEND- and clustering-based approaches
\cite{EEND-vector-clustering_ICASSP2021, EEND-vector-clustering_Interspeech2021,Horiguchi_ASRU2021, coria2021overlapaware},
referred to as EEND-vector clustering (EEND-VC) hereafter,
has been recently proposed to cope with realistic recordings containing overlapped speech
with an arbitrarily large number of speakers.
It first splits the input recording into fixed-length chunks.
Then, it applies a modified version of EEND to each chunk
to obtain diarization results for speakers speaking in each chunk
as well as speaker embeddings for them. 
Finally, to estimate which of the diarization results estimated in local chunks 
belongs to the same speaker, 
speaker clustering is performed across the chunks 
based on the speaker embeddings by using a constrained clustering algorithm.
While this integrated approach is shown to achieve state-of-the-art results
for real conversational data such as CALLHOME data \cite{EEND-vector-clustering_ICASSP2021,Horiguchi_ASRU2021},
we argue that there is a large room for improvement 
because the integration is not {\it tight} enough;
Although the estimation of diarization results and speaker embeddings
is based on a single NN and thus are tightly coupled,
the clustering stage is formulated as an independent process
that is not guaranteed to be optimal in clustering the speaker embeddings,
and thus the overall system could not be optimal.

To address this problem and tightly integrate EEND- and clustering-based diarization,
this paper introduces a trainable clustering framework,
unfolded infinite Gaussian mixture model (iGMM) \cite{iwata2021metalearning}, 
into the EEND-VC framework.
Desired properties of a clustering algorithm for EEND-VC are 
(1) it should deal with arbitrary unbounded number of speakers,
(2) it should estimate the number of speakers in an optimal sense,
(3) it should handle non-sequential data (unlike \cite{Zhang_ICASSP19}) because a set of the speaker embeddings
in the EEND-VC framework has no specific order.
As a typical clustering algorithm that fulfills these conditions,
we propose to employ a non-parametric Bayesian model called iGMM,
which is a GMM but with a theoretically infinite number of mixture components.
The number of mixture components, corresponding to the number of speakers 
in diarization, can be optimized in a maximum marginal likelihood sense, given an observation.
To jointly optimize this novel clustering step with
speaker embedding estimation and diarization results estimation,
we opt to {\it unfold} the parameter estimation process of iGMM
and optimize directly the clustering results 
through a novel adjusted Rand index (ARI)-based loss \cite{iwata2021metalearning}.
Experiments based on CALLHOME data show the proposed approach can outperform the conventional EEND-VC
in terms of diarization error rate (DER) especially by reducing speaker confusion errors,
which indeed reflects the effectiveness of the proposed iGMM integration.

% The remainder of this paper is organized as follows. We first introduce the proposed framework in section 2 in detail.
% Then, in section 3, we evaluate its performance in comparison with the original EEND to clarify the advantages of the proposed framework.
% Finally, we conclude the paper in section 4.

%\section{Related Work}

\begin{figure}[t]
 \begin{center}
  \includegraphics[width=90mm]{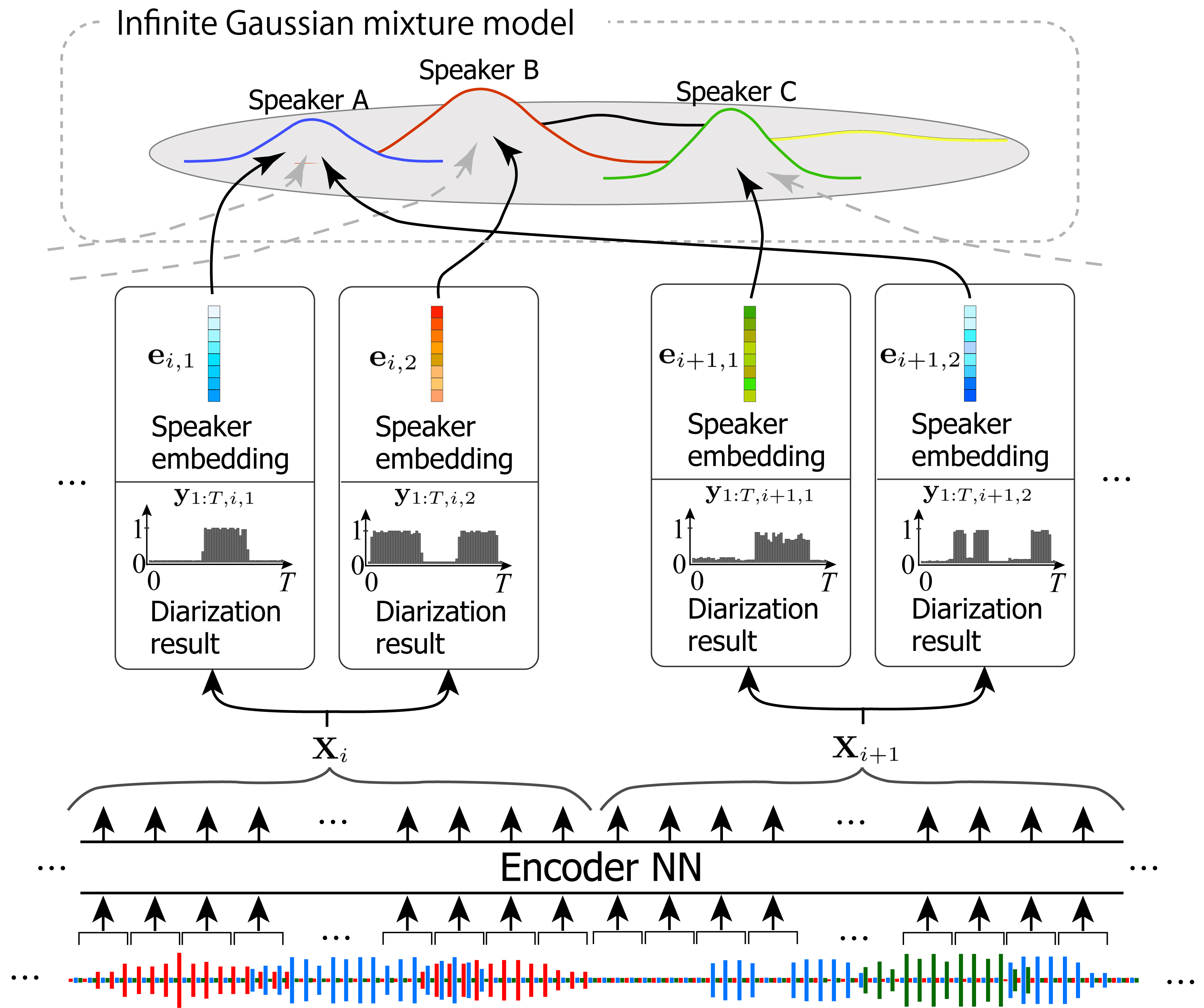}
     \end{center}
   \vspace{-3mm}
   \caption{Schematic diagram of the proposed diarization framework. The input contains 3 speakers in total (red, green, and blue speakers shown in the waveform at the bottom), but only at most 2 speakers are actively speaking in each chunk.}
 \label{fig:overview}
\end{figure}

\begin{figure}[t]
 \begin{center}
  \includegraphics[width=55mm]{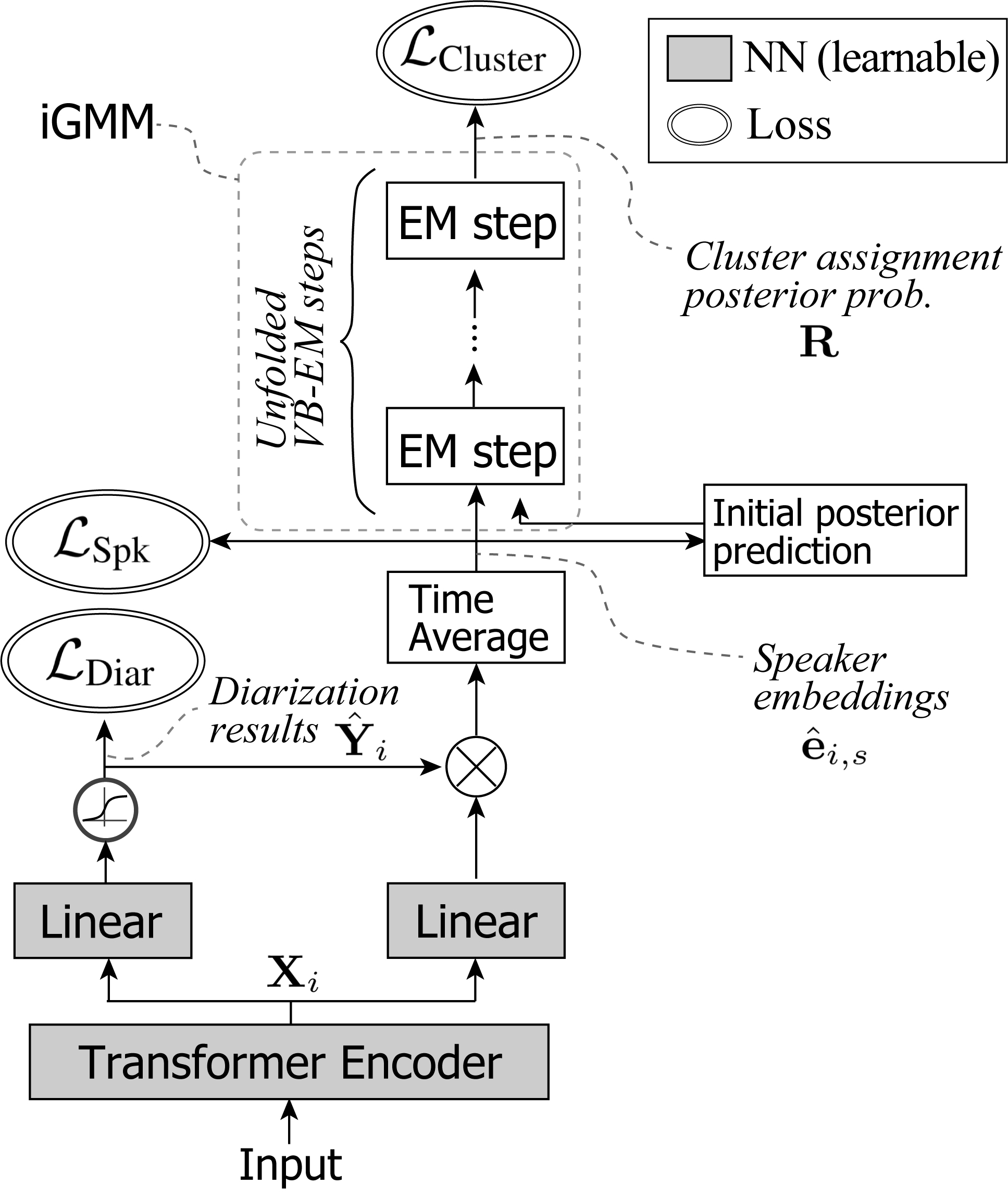}
     \end{center}
   \vspace{-4mm}
   \caption{Proposed neural network architecture and loss functions for neural network optimization}
 \label{fig:architecture}
\end{figure}

\section{Proposed Diarization framework}
% \vspace{-2mm}
% \label{sec:proposed_method}
% This section describes the proposed diarization framework in details,
% and clarify its advantages over conventional approaches.

\vspace{-2mm}
\subsection{Overall framework}
\vspace{-2mm}
\label{sec:framework}
Figure~\ref{fig:overview} shows a schematic diagram of the proposed framework, EEND-vector clustering with iGMM (EEND-VC-iGMM), for an exemplary 2 chunks out of continuous 3-speaker meeting data.

It first passes a several-minute long input recording to NN (``Encoder NN'' in Fig.~\ref{fig:overview}), 
and obtain a set of $D$-dimensional frame features.
Then, these features are segmented into chunks 
to form chunk-level features, 
as $\mathbf{X}_i = (\mathbf{x}_{t,i} \in \mathbb{R}^{D} \mid t=1,\cdots,T)$ where $i$, $t$ and $T$ are the chunk index, the frame index in the chunk and the chunk size.
%Fig~\ref{fig:overview} shows EEND-VC-iGMM processing for an exemplary 2 chunks out of 3-speaker meeting data. 
In the following explanation, we assume that we can reasonably fix the maximum number of active speakers in a chunk, $S_{\textrm{Local}}$, to 2, for the sake of simplicity
\footnote{In our experiments, the chunk size $T$ and $S_{\textrm{Local}}$ are set at 5~s
and 3, respectively.}. 

With the assumption/hyper-parameter $S_{\textrm{Local}}=2$ and $\mathbf{X}_i$,
the system estimates, based on an NN, 
diarization results and speaker embeddings associated with the 2 speakers in each chunk.
If a speaker is absent (i.e., there is only one active speaker in that chunk), 
the network simply estimates the diarization results of all zeros for that silent speaker.
Since it is {\it not} always guaranteed that the diarization results of a certain speaker are estimated at the same output node,
we may have the inter-chunk label permutation problem in the diarization outputs \cite{EEND-vector-clustering_ICASSP2021,Xue2020_speaker_tracing}.
We can solve this permutation problem and estimate the correct association of the diarization results among chunks, 
by clustering the speaker embeddings given the total number of speakers in the input recording, $S_{\textrm{Global}}$, (3 in the example shown in Fig.~\ref{fig:overview}), or given an estimate of $S_{\textrm{Global}}$.

In the previous studies \cite{EEND-vector-clustering_Interspeech2021,Horiguchi_ASRU2021},
the speaker embedding extraction process is optimized 
such that vectors of the same speaker stay close to each other,
while those from different speakers lie far away from each other,
based on a categorical cross-entropy loss \cite{EEND-vector-clustering_Interspeech2021}
or a contrastive loss \cite{Horiguchi_ASRU2021}.
Then, the obtained embeddings are clustered into each speaker,
utilizing constrained clustering algorithms 
such as constrained Agglomerative Hierarchical Clustering (AHC).
In other words, the embeddings were not optimized for clustering.

In this paper,
we propose to train speaker embeddings
such that they can be well modeled by a non-parametric Bayesian model called iGMM.
By having this clustering process not only in the inference but also in the training,
we can tightly integrate the speaker embedding estimation and the subsequent clustering processes.

In the next subsections, we will detail essential components of EEND-VC-iGMM
by using Fig.~\ref{fig:architecture},
which summarizes an NN processing flow and loss functions used in the proposed framework.

\subsection{Chunk-wise diarization and speaker embedding estimation}
\vspace{-2mm}
The lower part of Fig.~\ref{fig:architecture} corresponds to the chunk-wise estimation of
diarization results and speaker embeddings.
%$\mathrm{NN}$ in Fig.~\ref{fig:overview} can be formulated as follows. 
First let us denote the ground-truth diarization labels at each $i$-th chunk 
as $\mathbf{Y}_i = \{\mathbf{y}_{t,i} \mid t=1,\cdots,T \} \in \mathbb{R}^{S_{\textrm{Local}} \times T}$ that corresponds to $\mathbf{X}_i$,
and $C$-dimensional speaker embeddings estimated at the $i$-th chunk for the $s$-th speaker as 
$\hat{\mathbf{e}}_{i,s} \in \mathbb{R}^{C}$.
Here, the diarization label $\mathbf{y}_{t,i} = [y_{t,i,s} \in \{0,1\} \mid s=1, \cdots, S_{\textrm{Local}}]$ represents a joint activity for $S_{\textrm{Local}}$ speakers.
For example, $y_{t,i,s} = y_{t,i,s'} = 1 (s \ne s')$ indicates both speakers $s$ and $s'$ spoke at the time frame $t$ in the chunk $i$. 
Then, after obtaining $\{\mathbf{X}_i\}_{i=1}^{I}$ for all $I$ chunks
by processing input speech with a Transformer encoder,
we can jointly estimate diarization results and speaker embeddings at the $i$-th chunk as:
\begin{align}
\vspace{-1mm}
    \hat{\mathbf{Y}}_i = \sigma(\mathrm{Linear}( \mathbf{X}_i )), \ \ 
    \hat{\mathbf{e}}_{i,1},\ldots,\hat{\mathbf{e}}_{i,S_{\textrm{Local}}} = \mathrm{Avg}_{\hat{\mathbf{Y}}_i}(\mathrm{Linear}( \mathbf{X}_i )). \nonumber %\label{eq:NN_out}
    %\hat{\mathbf{E}}_i = \mathrm{Linear}( \mathbf{X}_i ), \nonumber %\label{eq:NN_out}
\vspace{-1mm}
\end{align}
$\sigma(\cdot)$, $\mathrm{Linear}(\cdot)$ and $\mathrm{Avg}_{\mathbf{A}}(\cdot)$ are the sigmoid activation function, linear layers, and a time-averaging function with a time-varying weight $\mathbf{A}$,
respectively.

% $\mathrm{NN}$ can be trained with a multitask loss function composed of diarization loss, i.e., binary cross entropy loss,
% and speaker embedding loss that encourages the embeddings to have small intra-speaker and large inter-speaker euclidean distances,
% as proposed in \cite{EEND-vector-clustering_ICASSP2021}.

\subsection{Infinite Gaussian mixture model and its deep unfolding}
\vspace{-2mm}
After estimating a set of speaker embeddings $\{\{\hat{\mathbf{e}}_{i,s}\}_{s=1}^{S_{\textrm{Local}}}\}_{i=1}^{I}$, 
we cluster them with an iGMM, which is a special case of Dirichlet process (DP) mixture models.
The upper right part of Fig.~\ref{fig:architecture} corresponds to the speaker embedding clustering process by iGMM.
iGMM theoretically has an infinite number of Gaussian components,
and uses a part of them to appropriately model observation data.
The clustering and the number of cluster estimation are jointly done 
in a maximum marginal likelihood sense by means of variational Bayesian (VB) inference 
of the model parameters given observed data.
% It is typically used in cases where we don't know the number of clusters in advance.
% This model fits perfectly the EEND-vector clustering framework,
% because the number of speakers in an input recording is not known in advance,
% is generally unbounded, and has to be estimated jointly with the clustering process. 

Specifically, the proposed iGMM takes a set of speaker embeddings as an input,
and outputs soft speaker-cluster assignments 
$\mathbf{R}=\{\{\mathbf{r}_{i,s}\}_{s=1}^{S_{\textrm{Local}}}\}_{i=1}^{I}$,
where $\mathbf{r}_{i,s} =\{r_{i,s,k}\}_{k=1}^{K'}$.
$r_{i,s,k}$ is the probability that the speaker embedding $\hat{\mathbf{e}}_{i,s}$ is assigned to the $k$-th cluster,
% $0 \leq \hat{r}_{i,s,k}\leq 1$, $\sum_{k=1}^{K'} \hat{r}_{i,s,k}=1$,
and $K'$ is the maximum number of clusters, which is set at a large value in practice.

\subsubsection{Generative process of the speaker embeddings}
\vspace{-2mm}
First, let us explain the generative process assumed in the proposed iGMM.
For the sake of convenience, let us introduce a variable $N$ that corresponds
to the total number of input speaker embeddings, i.e., $N= I \times S_{\textrm{Local}}$,
and an index for the embeddings, $n$, such as $\mathbf{e}_{n}$
\footnote{This index conversion is possible because the obtained speaker embeddings are non-sequential data.}.
%Given the input, we update cluster assignments by fitting an iGMM based on the VB inference.
% \footnote{Since the set of speaker embeddings are non-sequential data, we can }.
Then, in this paper, we employ a spherical iGMM with the following generative process for the speaker embeddings,
where the mixture weights are constructed by a DP prior
with concentration parameter $\alpha$ by a stick-breaking process~\cite{sethuraman1994constructive}, as:  
\begin{enumerate}
\item For each speaker cluster $k=1,\cdots,\infty$\vspace{-1mm}
  \begin{enumerate}
  \item Draw stick proportion $\eta_{k}\sim\mathrm{Beta}(1,\alpha)$ \vspace{-1mm}
  \item Set mixture weight $\pi_{k}=\eta_{k}\prod_{k'=1}^{k-1}(1-\eta_{k'})$\vspace{-1mm}
  \item Draw cluster mean $\bm{\mu}_{k}\sim\mathcal{N}(\mathbf{0},\mathbf{I})$\vspace{-1mm}
  \item Draw cluster precision $\beta_{k}\sim\mathrm{Gamma}(1,1)$\vspace{-1mm}
  \end{enumerate}
\item For each speaker embedding $n=1,\cdots,N$\vspace{-1mm}
  \begin{enumerate}
  \item Draw cluster assignment\\ $v_{n}\sim \mathrm{Categorical}(\bm{\pi})$\vspace{-1mm}
  \item Draw instance representation\\ $\mathbf{e}_{n}=\mathcal{N}(\bm{\mu}_{v_{n}},\beta_{v_{n}}^{-1}\mathbf{I})$\vspace{-0mm}
  \end{enumerate}
\end{enumerate}
$\mathrm{Beta}$ is the beta distribution,
$\mathcal{N}(\bm{\mu},\bm{\Sigma})$ is the Gaussian distribution with mean $\bm{\mu}$ and covariance $\bm{\Sigma}$,
$\mathrm{Gamma}$ is the gamma distribution,
$\mathrm{Categorical}$ is the categorical distribution, and
$\bm{\pi}=\{\pi_{k}\}_{k=1}^{\infty}$.
The DP prior by a stick-breaking process (steps 1-(a) and 1-(b)) is a key to allow us to use the infinite number of mixture components.

\vspace{-2mm}
\subsubsection{Parameter estimation for iGMM}
\label{sec:EM}
\vspace{-2mm}
Following the above generative process, 
we can derive the following parameter estimation steps based on the VB expectation-maximization (EM) algorithm.
Because of the space limitation, the derivation of the following equations is omitted, 
but it follows a straight-forward procedure of maximizing an evidence lower bound derived from the iGMM likelihood
as shown in \cite{iwata2021metalearning}.
The iGMM parameter estimation in the variational posterior distributions
is achieved by alternately calculating the following VB M-step:
\begin{align}
\vspace{-2mm}
  \gamma_{k1}&=1+\sum_{n=1}^{N}r_{n,k},\quad
  \gamma_{k2}=\alpha+\sum_{n=1}^{N}\sum_{k'=k+1}^{K'}r_{n,k'},\quad \nonumber \\
  \bm{\theta}_{k} &= \frac{\frac{b_{k}}{a_{k}}\sum_{n=1}^{N}r_{n,k}\mathbf{e}_{n}}
     {1+\frac{b_{k}}{a_{k}}\sum_{n=1}^{N}r_{n,k}},\nonumber \\ 
     a_{k}&=1+\frac{C}{2}\sum_{n=1}^{N}r_{n,k},\quad
  b_{k}=1+\frac{1}{2}\sum_{n=1}^{N}r_{n,k}(\parallel\mathbf{e}_{n}-\bm{\theta}_{k}\parallel^{2}+C), \nonumber
\end{align}
\vspace{-1mm}
and the following VB E-step to obtain a cluster assignment $r_{n,k}$:
\begin{align}
\vspace{-2mm}
  & \log r_{n,k} \propto
  \Psi(\gamma_{k1})-\Psi(\gamma_{k1}+\gamma_{k2})
  -\frac{C}{2}(\Psi(a_{k})-\log(b_{k})) \nonumber \\%\label{eq:r}  \\
  &-\frac{a_{k}}{2b_{k}}(\parallel\mathbf{e}_{n}-\bm{\theta}_{k}\parallel^{2}+C)
 %\nonumber\\
  +\sum_{k'=k+1}^{K'}(\Psi(\gamma_{k2})-\Psi(\gamma_{k1}+\gamma_{k2})), \nonumber
  \vspace{-1mm}
\end{align}
where $\Psi$ is the digamma function.
For the computational efficiency,
we truncate the number of clusters at $K'$ as in~\cite{blei2004variational}.
Note that the truncated DP is shown to closely approximate a true
DP for large enough $K'$ relative to the number of samples~\cite{ishwaran2001gibbs}.

Since \cite{iwata2021metalearning} shows that,
to help the VB EM steps converge faster to a better solution,
it is beneficial to estimate an initial value of the posterior probability $\mathbf{R}$
with another small NN,
we also employ such a small network,
which corresponds to a block denoted as ``Initial posterior prediction'' in Fig.~\ref{fig:architecture}. Its details are summarized in \cite{iwata2021metalearning}.

\subsubsection{Deep unfolding of iGMM parameter estimation process}
\vspace{-2mm}
The above VB EM steps are all clearly differentiable.
Thus, by following an idea of general deep unfolding framework, e.g., \cite{hershey2014deep},
we unfold the EM iterations into a sequential processing as in the upper right part of Fig.~\ref{fig:architecture}
to incorporate iGMM-based clustering into the overall NN optimization framework.
% by unfolding the EM iterations as in the upper right part of Fig.~\ref{fig:architecture} 
% we can backpropagate the error in the posterior probability estimation to the input
% to fix how each speaker embedding should be formed, to be better fit to the iGMM-based clustering.
% This follows an idea of general deep unfolding framework, e.g., \cite{hershey2014deep}.
% In the actual processing, we interpret the EM iterations as this unfolded processing.

\subsection{Loss functions}
\vspace{-2mm}
Now, let us explain how we optimize the network.
As it is shown in Fig.~\ref{fig:architecture}, the system can be optimized by the following multi-task loss;
\begin{eqnarray}
\vspace{-1mm}
\mathcal{L} &=& (1-\lambda_{1}-\lambda_{2}) \mathcal{L}_{\textrm{Diar}} + \lambda_{1} \mathcal{L}_{\textrm{Cluster}} + \lambda_{2} \mathcal{L}_{\textrm{Spk}}, \label{eq:total_loss} 
\vspace{-1mm}
\end{eqnarray}
where $\mathcal{L}_{\textrm{Diar}}$, $\mathcal{L}_{\textrm{Cluster}}$, $\mathcal{L}_{\textrm{Spk}}$
correspond to losses that control chunk-wise diarization accuracy, clustering accuracy, 
and a speaker embedding space to have small intra-speaker and large inter-speaker variability, respectively. 
$\mathcal{W}=\{\lambda_{1}, \lambda_{2}\}$ includes weights for the multi-task loss.
In the following, we will detail 
$\mathcal{L}_{\textrm{Diar}}$ and $\mathcal{L}_{\textrm{Cluster}}$,
while we ask readers to refer to \cite{EEND-vector-clustering_ICASSP2021} for details of $\mathcal{L}_{\textrm{Spk}}$.
$\mathcal{L}_{\textrm{Spk}}$ in this paper is based on absolute speaker identity labels.

\begin{table*}[t]
\centering
\caption{DERs (\%) of EEND-VC and the proposed EEND-VC-iGMM for the different number of speakers in the CALLHOME evaluation set. The numbers in parentheses indicate the missed detection (MI), false alarm (FA) and speaker confusion (CF) errors, i.e., (MI/FA/CF).}
\vspace{-3mm}
\label{tab:results}
\scalebox{0.95}[0.95]{
\begin{tabular}{l@{ }c@{ }c c c c cc}
\toprule
\multirow{2}{*}{Model}     & \multirow{2}{*}{$\mathcal{L}_{\textrm{Spk}}$}     &   \multicolumn{6}{c}{Number of speakers in a recording}  \\ \cline{3-7}
                           &                            &   2 & 3     & 4 & 5 & 6 & Avg. \\ 
\midrule
EEND-VC   & \checkmark& \textbf{7.0} (4.0/2.4/0.5) &	14.2 (4.9/3.5/5.8)	&16.7 (6.0/2.7/8.0) &	31.6 (8.0/2.4/21.2)&	29.9 (10.9/3.5/15.5) & 13.8 (5.6/2.7/5.5)    \\
\midrule
EEND-VC-iGMM  &      -     &  8.1 (4.6/2.7/1.1)&	\textbf{12.3} (5.2/3.6/3.5)	&18.0 (6.4/4.4/7.1)&	28.4 (6.3/4.7/17.3)	&33.8 (11.1/4.8/17.9)	&13.7 (6.5/2.7/4.5)  \\ 
EEND-VC-iGMM  & \checkmark &     	8.6 (4.8/2.3/1.4) &	12.6 (6.6/2.3/3.6)	&\textbf{16.1} (6.4/3.7/6.1)	&\textbf{27.5} (5.3/4.9/17.3)&	\textbf{26.9} (11.9/3.2/11.4) &	\textbf{13.3} (5.2/3.6/4.5) \\
%EEND-VC-iGMM  & \checkmark &     	7.7(3.7,3.3,0.7)	&\textbf{12.1}(5.8,3.0,3.4)	&17.2(6.9,3.2,7.1)&	\textbf{29.2}(5.3,5.9,18.0)	&\textbf{28.8}(12.4,3.6,12.8)	&\textbf{13.2}(5.6,3.3,4.3)    \\ 
\bottomrule
\end{tabular}}
\end{table*}

\subsubsection{Chunk-level diarization loss}
\vspace{-2mm}
As in \cite{Fujita_IS2019}, the diariation loss $\mathcal{L}_{\textrm{Diar}}$ in each chunk is formulated as:
\begin{eqnarray}
\vspace{-2mm}
\mathcal{L}_{\textrm{Diar},i} &=& \frac{1}{TS_{\textrm{Local}}} \min_{\phi \in \mathrm{perm}(S_{\textrm{Local}})} \sum_{t=1}^{T} \textrm{BCE}\left( \mathbf{l}_{t,i}^{\phi},\hat{\mathbf{y}}_{t,i}  \right), \label{eq:diarization_loss} 
\vspace{-1mm}
\end{eqnarray}
where $\mathrm{perm}(S_{\textrm{Local}})$ is the set of all the possible permutations of ($1,\dots,S_{\textrm{Local}}$),
$\hat{\mathbf{y}}_{t,i}=[\hat{y}_{t,i,1},\ldots,\hat{y}_{t,i,S_{\textrm{Local}}}] \in \mathbb{R}^{S_{\textrm{Local}}}$,
$\mathbf{l}_{t,i}^{\phi}$ is the $\phi$-th permutation of the reference speaker labels, and $\mathrm{BCE}(\cdot, \cdot)$ is the binary cross-entropy function between the labels and the estimated diarization outputs.
% $\phi^{\star}$ is the permutation that minimizes the right hand side of the Eq.~(\ref{eq:diarization_loss}).

\subsubsection{Clustering loss: Adjusted Rand index loss}
\vspace{-2mm}
A common practice to evaluate clustering accuracy is to use the ARI
\cite{rand1971objective,hubert1985comparing,vinh2010information},
that directly measures similarity between a ground-truth clustering result and an estimated one,
even when the estimated and true number of clusters does not agree.
We here propose to use the (negative) ARI as a loss to directly improve the accuracy of the iGMM-based speaker embedding clustering, 
i.e., the accuracy of the posterior probability $\mathbf{R}$ obtained in \ref{sec:EM}.

Specifically, we use the following continuous approximation of ARI~\cite{iwata2021metalearning} (hereafter, cARI)
that can handle soft cluster assignments, as opposed to the original non-differentiable ARI.
Let us first define 
$N_{1}$ as the approximated number of pairs of instances (i.e., speaker embeddings) that are in different clusters in both the true and estimated assignments,
$N_{2}$ as the approximated number of pairs that are in different clusters in the true assignments but not in the estimated assignments,
$N_{3}$ as the approximated number of pairs that are in the same cluster in the true assignments but not in the estimated assignments, and
$N_{4}$ as the approximated number of pairs that are in the same cluster in both the true and estimated assignments.
Then, cARI is formulated as:
\begin{align}
  \mathrm{cARI}=\frac{2(N_{1}N_{4}-N_{2}N_{3})}{(N_{1}+N_{2})(N_{3}+N_{4})+(N_{1}+N_{3})(N_{2}+N_{4})},
  \label{eq:smoothed_ari}
\end{align}
where $N_i (i=1,\ldots,4)$ is mathematically defined as:
\vspace{-1mm}
\begin{align}
\vspace{-1mm}
  N_{1}&=\sum_{n=1}^{N}\sum_{n'=n+1}^{N}I(h_{n}\neq h_{n'})d_{n,n'}, \vspace{-1mm}\\
  N_{2}&=\sum_{n=1}^{N}\sum_{n'=n+1}^{N}I(h_{n}\neq h_{n'})(1-d_{n,n'}),\vspace{-1mm}
\end{align}
\begin{align}
  N_{3}&=\sum_{n=1}^{N}\sum_{n'=n+1}^{N}I(h_{n}=h_{n'})d_{n,n'},\vspace{-1mm} \\
  N_{4}&=\sum_{n=1}^{N}\sum_{n'=n+1}^{N}I(h_{n}=h_{n'})(1-d_{n,n'}),\vspace{-1mm}
\end{align}
where $h_{n}$ is the true cluster assignment label for the $n$-th speaker embedding $\mathbf{e}_{n}$, 
$I(\cdot)$ is the indicator function, i.e., $I(A)$=1 if $A$ is true and $0$ otherwise,
and $d_{n,n'}$ is the total variation distance~\cite{gibbs2002choosing} between $\mathbf{r}_{n}$ and $\mathbf{r}_{n'}$
defined as:
\begin{align}
\vspace{-1mm}
  d_{n,n'}=\frac{1}{2}\sum_{k=1}^{K'}|r_{n,k}-r_{n',k}|.
  \label{eq:d}
\vspace{-1mm}
\end{align}
As a loss function $\mathcal{L}_{\textrm{Cluster}}$, we minimize $\mathcal{L}_{\textrm{Cluster}}=-\mathrm{cARI}$.

\vspace{-0mm}
\section{Experiments}
\vspace{-2mm}
Here, we evaluate the effectiveness of the proposed EEND-VC-iGMM on the widely used CALLHOME (CH) dataset \cite{CALLHOME,Diarization_review}. 

\vspace{-2mm}
\subsection{Data}
\vspace{-2mm}
We trained the diarization systems on simulated mixtures using speech from Switchboard-2, 
Switchboard Cellular, and the NIST Speaker Recognition Evaluations,
and noise from the MUSAN corpus \cite{MUSAN}, and simulated room impulse responses from \cite{Ko_2017}.

We generated 2 sets of training data. The first set (6.9k~hours) consists of 1-to-3-speaker meeting-like data generated 
based on the algorithm proposed in \cite{Fujita_IS2019} with $\beta=10$. 
% \footnote{Each mixture contains dozens of utterances per speaker with reasonable silence intervals between utterances of the same speaker's. 
% $\beta=10$ means that the average duration of the silence interval is 10~s.
% }. 
This is the same training dataset as the one we used in \cite{EEND-vector-clustering_Interspeech2021}. 
We used it to train a seed model that was common for our baseline and proposed systems. 
The second training data (5.5k~hours) consists of mixtures of up to 7 speakers, 
which simulates meetings with a larger number of speakers. 
% The amount of training data was 6900 hours for the first training set and 5500 hours for the second.

We evaluated the diarization systems on the CH dataset  \cite{CALLHOME} 
that contains 500 telephone-conversation sessions including 2 to 6 speakers. 
Because there is a mismatch between the training and testing conditions, 
we used a part of the CH data for adaptation. 
We use the adaptation/evaluation data split proposed in \cite{Horiguchi2020_EDA_EEND}.
% and performed adaptation on a subset and evaluation of the proposed method on the other subset.

\subsection{Experimental settings}
\vspace{-2mm}
We evaluate the proposed EEND-VC-iGMM 
in comparison with the original EEND-VC with constrained AHC 
\cite{EEND-vector-clustering_Interspeech2021}. 
Both systems use the same configuration for the input feature,
the Transformer encoder network, and the same {\it silent speaker} detection, all of which follow \cite{EEND-vector-clustering_Interspeech2021}.
The only difference comes from the clustering modules, i.e., constrained AHC versus trainable iGMM. 
% The propose EEND-VC-iGMM has an additional one linear layer
% to convert the speaker embedding vectors before iGMM clustering. 
We assume a maximum number of speakers per chunk to be 3,
i.e., $S_{\mathrm{local}}=3$.
% Consequently, the diarization systems always estimate 3 diarization outputs 
% and 3 speaker embeddings including silent speakers.
NN of the EEND-VC was trained with the multi-task weight of 
$\mathcal{W}=\{0.0,0.03\}$.
We prepared two variants of EEND-VC-iGMM, one with the speaker embedding loss $\mathcal{L}_{\textrm{Spk}}$ based on absolute
speaker identify labels, 
and one without it,
by setting $\mathcal{W}=\{0.05,0.03\}$
and $\mathcal{W}=\{0.05,0.0\}$, respectively.
% In all systems, before clustering speaker embeddings,
% embeddings corresponding to {\it silent speakers} are automatically detected and excluded from clustering, as in \cite{EEND-vector-clustering_Interspeech2021}.

The training procedure is as follows. We first created the seed model using the 1-to-3-speaker training data and 30 seconds chunks for 100 epochs. We then re-trained the baseline EEND-VC and proposed EEND-VC-iGMM on the 2-to-7-speaker training data with 5-second chunks, i.e., $T=5~s$. These chunks are taken from 100~s and 300~s consecutive recordings for EEND-VC and EEND-VC-iGMM, respectively.  
Finally, we performed adaption using the CALLHOME adaptation data. 
For adaptation, we cut the recordings to 100~s for the baseline and 600~s for the proposed method, which corresponds to the optimal setting for each. This setting allows EEND-VC-iGMM to have sufficient number of embedding samples for the iGMM clustering during training.
For the iGMM, we set the number of EM iterations at 10, $\alpha$ at 1, $K'$ at 10, in both training and inference stages.

The performance was evaluated including overlapped speech frames in terms of DER with a collar tolerance of 0.25~s as in \cite{Horiguchi2020_EDA_EEND,Diarization_review}.

\vspace{-2mm}
\subsection{Results}
\vspace{-2mm}
Table \ref{tab:results} shows the DERs for the conventional EEND-VC, and EEND-VC-iGMM with and without 
the speaker embedding loss $\mathcal{L}_{\textrm{Spk}}$.
We can see that EEND-VC-iGMM outperforms EEND-VC in all but the 2-speaker condition. 
By looking at the breakdown of DERs, 
we observe that EEND-VC-iGMM greatly reduces speaker confusion errors in most cases. 
This clearly confirms the effectiveness of incorporating the trainable iGMM-based clustering and tightly coupling 
the embedding estimation and the clustering stages.

Looking at Avg. conditions, we can see that EEND-VC-iGMM with the speaker embedding loss $\mathcal{L}_{\textrm{Spk}}$ performed the best.
Another variant of EEND-VC-iGMM that does not use $\mathcal{L}_{\textrm{Spk}}$, which is based on absolute speaker identity
labels, achieves overall performance comparable to the baseline but with lower speaker confusion errors.
Unlike the baseline, this proposed variant does not require 
absolute speaker identity labels and relies only on diarization labels. 
Considering that (1) the performance of EEND-VC is fairly
good on this data in general and (2) there are many cases that 
the absolute speaker identify labels are not available,
this is an encouraging result.

% Since absolute speaker labels are not always easily available, EEND-VC-iGMM offers a promising alternative.
% The performance of EEND-VC-iGMM that does not use $\mathcal{L}_{\textrm{Spk}}$ based on absolute speaker identity
% label is comparable with the baseline EEND-VC that 
% relies on the $\mathcal{L}_{\textrm{Spk}}$.
% Considering that (1) the performance of EEND-VC is fairly
% good on this data in general and (2) there are many cases that 
% the absolute speaker identify labels are not available,
% this is an encouraging result.
% Note that even though their DERs are comaparable,
% we can still see that the speaker confusion errors are 
% much lower for EEND-VC-iGMM.

The numbers reported in Table~\ref{tab:results} are slightly worse than those reported in \cite{EEND-vector-clustering_Interspeech2021}, because of the different chunk size (i.e., we use here a chunk size $T$ of 5~s, while the best performance in \cite{EEND-vector-clustering_Interspeech2021} was achieved with a chunk size of 30~s). 
Although the chunk size of 5~s may not be optimal for the CH data, it is arguably a much more practical setting in general as it allows us to cope with conversations with rapid speaker changes such as a meeting or casual conversations.
In future work, we plan to investigate the proposed EEND-VC-iGMM in such challenging conditions.

\vspace{-2mm}
\section{Conclusion}
\vspace{-2mm}
This paper introduced a trainable clustering, i.e., deep unfolded iGMM, into the EEND-VC framework,
that allows tighter integration of EEND-based and clustering-based diarization approaches.
We confirmed experimentally that the proposed method could outperform the conventional EEND-VC with constrained AHC, 
by significantly reducing the speaker confusion errors. 
% Our future works will include investigating the method on more challenging conditions such as CHiME 6[??].

\bibliographystyle{IEEEbib}
\bibliography{strings,mybib}

\end{document}